\documentclass[referee]{aa}
\usepackage[varg]{txfonts}
\usepackage{graphicx}
\usepackage{epstopdf}
\usepackage{mathtools}
\usepackage{multirow}
\usepackage{natbib}
\begin{document}

\title{Analysis of Moon impact flashes detected during the 2012 and 2013 Perseids}

\author{José M. Madiedo\inst{1,2}
  \and José L. Ortiz\inst{3} 
  \and Faustino Organero\inst{4}
  \and Leonor Ana-Hernández\inst{4}  
  \and Fernando Fonseca\inst{4}
  \and Nicolás Morales\inst{3} 
  \and Jesús Cabrera-Caño\inst{1}}

\institute{Departamento de Física Atómica, Molecular y Nuclear. Facultad de Física. Universidad de Sevilla. 41012 Sevilla, Spain
	\and Facultad de Ciencias Experimentales, Universidad de Huelva, 21071 Huelva, Spain 
  \and Instituto de Astrofísica de Andalucía, CSIC, Apt. 3004, 18080 Granada, Spain
  \and Observatorio Astronómico de La Hita, La Puebla de Almoradiel, Toledo, Spain} 

\date{Received 15 January 2015 / Accepted 11 March 2015}

\abstract{We present the results of our Moon impact flashes detection campaigns performed around the maximum activity period of the Perseid meteor shower in 2012 and 2013. Just one flash produced by a Perseid meteoroid was detected in 2012 because of very unfavourable geometric conditions, but 12 of these were confirmed in 2013. The visual magnitude of the flashes ranged between 6.6 and 9.3. A luminous efficiency of 1.8·10$^{-3}$ has been estimated for meteoroids from this stream. According to this value, impactor masses would range between 1.9 and 190 g. In addition, we propose a criterion to establish, from a statistical point of view, the likely origin of impact flashes recorded on the lunar surface.} 

\titlerunning{Moon impact flashes during the 2012 and 2013 Perseids}
\authorrunning{José M. Madiedo et al.}
\maketitle
\section{Introduction}
One of the techniques suitable to analyze the flux of meteoroids impacting the Earth is based on the monitoring of the night side of the Moon in order to detect flashes produced by the impact of these particles of interplanetary matter (Ortiz et al. 1999; Bellot Rubio et al. 2000a,b; Ortiz et al. 2000; Yanagisawa and Kisaichi 2002; Cudnik et al. 2002; Ortiz et al. 2002; Yanagisawa et al. 2006; Cooke et al. 2006; Ortiz et al. 2006; Madiedo et al. 2013). This method has the advantage that a single detector covers a much larger area (typically $\sim$ 10$^{6}$ km$^{2}$) than that monitored by ground-based systems that analyze meteor and fireball activity in the Earth’s atmosphere. However, the results derived from the analysis of lunar impact flashes depend strongly on the value adopted for the so-called luminous efficiency, i.e., the fraction of the kinetic energy of the impactor that is emitted as visible light during the impact. These results include, for instance, meteoroid mass, crater diameter and impact flux.

Experiments suggest that the luminous efficiency would depend on impact velocity, which means that the value of this parameter is stream dependent (Swift et al. 2011). So, a systematic monitoring of moon impact flashes would be desirable in order to obtain more precise values for the luminous efficiency for meteoroids from different sources. In this context, our team is performing a monitoring of the night side of the Moon from our observatories in the south and the center of Spain, two regions which provide very favourable statistics of clear nights per year. For this purpose we employ several telescopes equipped with high-sensitivity CCD video cameras, and we have also developed our own software to identify and analyze impact flashes. This research is conducted in the framework of a project started in 2009 and named MIDAS, which is the acronym for Moon Impacts Detection and Analysis System (Madiedo et al. 2010).

Since the cross section of meteoroid streams is larger than the Earth-Moon distance, the chances for impact flashes detection is higher during the activity period of major meteor showers, such as for instance the Perseids. The first confirmed lunar impact flash produced by a meteoroid belonging to the Perseid meteoroid stream was recorded and analyzed by Yanagisawa et al. (2006). The analysis of this single event suggested that the luminous efficiency for Perseid meteoroids would be of about 2.1·10$^{-4}$, which corresponds to the lower limit found for this parameter from the observation of Leonid lunar impact flashes (Bellot Rubio et al. 2000a,b). However, the most likely value found by Bellot Rubio et al. was of about 2·10$^{-3}$, which is also the value employed by Ortiz et al. (2006) to determine impact fluxes on Earth, and very close to the 1.5·10$^{-3}$ value taken by other researchers (Swift et al. 2011, Bouley et al. 2012). So, additional observations and analysis of Perseid impact flashes would be desirable in order to better constraint the value of this parameter for meteoroids from this stream. Here we present the results derived from the Moon impact flashes monitoring campaigns developed by our team during the maximum activity period of the Perseid meteor shower in 2012 and 2013.  We focus on the estimation of the luminous efficiency for Perseid meteoroids. On the other hand, we also address another issue related to the lunar impact flashes technique. Thus, in contrast to the method based on the observation and analysis of meteors and fireballs in the atmosphere, it is not possible to unambiguously establish the source of the impact events detected by means of this method. In this work we propose a criterion that allows quantifying the probability that an impact flash can be associated to a given meteoroid source.
\section{Instrumentation and methods}
\begin{figure*}
\resizebox{\hsize}{!}{\includegraphics{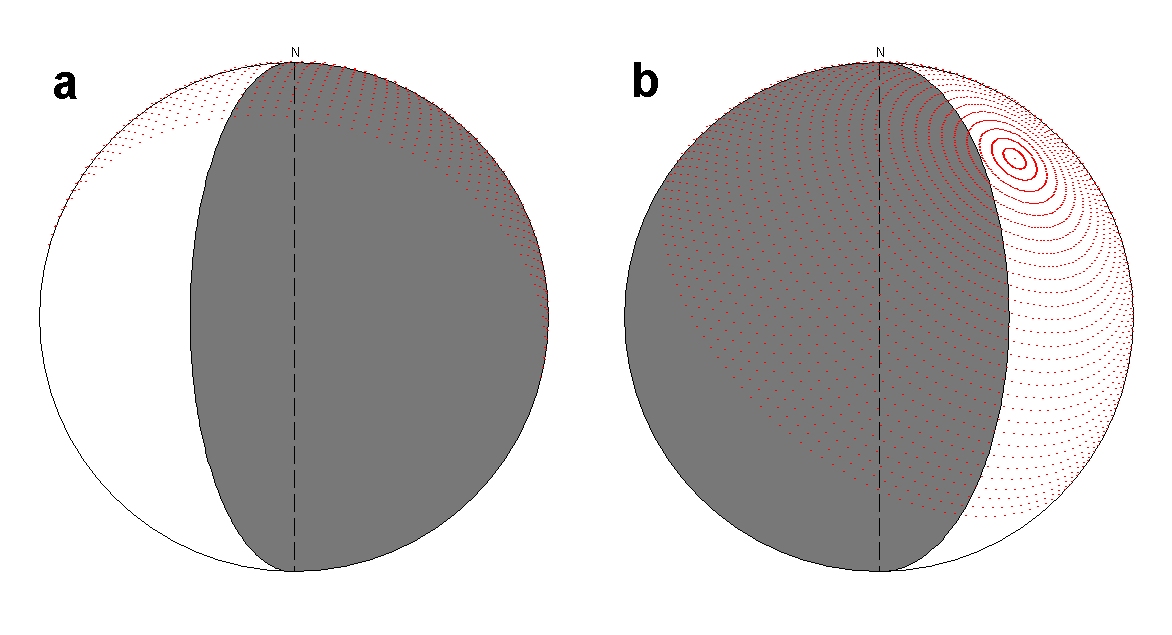}}  
  \caption{The lunar disk on 2012 August 12 (a) and 2013 August 12 (b). The white region corresponds to the area illuminated by the Sun. The gray region is the night side, as seen from our planet. The dotted region delimitates the area where Perseid meteoroids could impact.}
  \label{figure:1}
\end{figure*}
Our lunar impact flashes monitoring project is being developed from two different locations in Spain. Thus, from our observatory in Sevilla (latitude: 37.34611 ºN, longitude: 5.98055 ºW, height: 23 m above the sea level) we have operated two identical 0.36 m Schmidt-Cassegrain telescopes, but also a smaller Schmidt-Cassegrain telescope with a diameter of 0.28 m. All of them are manufactured by Celestron. On the other hand since 2013, a 40 cm Newtonian telescope is operated at La Hita Astronomical Observatory, in central Spain (latitude: 39.56833 ºN, longitude: 3.18333 ºW, height: 674 m above the sea level).

Each telescope employs a high-sensitivity CCD video camera (model 902H Ultimate, manufactured by Watec Corporation). These devices employ a Sony ICX439ALL 1/2" monochrome CCD sensor and produce interlaced analogue imagery according to the PAL video standard. Thus, images are obtained with a resolution of 720x576 pixels and a frame rate of 25 frames per second (fps). GPS time inserters are used to stamp time information on every video frame with an accuracy of 0.01 s. Besides, f/3.3 focal reducers manufactured by Meade are also used in order to increase the area monitored by these devices. The telescopes are tracked at lunar rate, but they are manually recentered when necessary, since perfect tracking of the Moon at the required precision is not feasible with this equipment.

The telescopes located at Sevilla are operated manually. The system at La Hita, however, is configured to work remotely, although it can be also operated in situ. When no major showers are active, these telescopes are oriented to an arbitrary region on the Moon surface in order to cover a common maximum area. However, as explained below, during the campaigns considered here we previously determined which areas on the lunar disk should be monitored in order to increase the possibility of Perseid flashes detection. With this information, the telescopes were aimed at the corresponding areas and the images were continuously recorded on multimedia hard disks. Of course, the terminator was avoided in order to avoid saturation of the CCD and also to prevent an excess of light from the illuminated side of the Moon in the telescopes.

Impact flashes are very short in duration (most of them are contained in just one or two video frames). So, their identification by hand is not practical and computer software is required to automatically identify impact candidates. With this aim we developed the MIDAS software, which in this context is the acronym for Moon Impacts Detection and Analysis Software (Madiedo et al. 2010). This package, which is also employed for data reduction and analysis of confirmed flashes, is described below.
\subsection{The MIDAS software}
The MIDAS software was developed with the aim to process live video streaming generated by the cameras attached to our telescopes, or previously recorded video files containing images of the night side of the Moon in order to automatically detect flashes produced by the impact of meteoroids on the lunar surface. Nevertheless, instead of performing a live analysis of the imagery obtained by this equipment, we prefer to store first the corresponding video files on a hard disk and process them later on. In addition, MIDAS contains several modules that can carry out different tasks. Most of these are related to data reduction, but others are employed to plan the observations. Thus, for instance, the software can determine which region on the lunar surface should be monitored to increase the chance of detecting events from a given meteoroid stream. For this purpose, the subradiant point on the Moon is calculated by following the method detailed in Bellot-Rubio et al. (2000b). One of the advantages of this software is its ability to perform different tasks simultaneously. Thus, for instance, while the impact flashes identification process is in progress, the user can view, edit, remove false detections and even analyze impact suspects detected by MIDAS.

In order to indentify an impact flash, the software compares consecutive video frames and detects brightness changes that exceed a given (user defined) threshold value. The selected detection settings are fundamental in order to get optimal results. These parameters depend, among other things, on the configuration of the imaging telescope and camera. The user must provide which is the minimum size and brightness threshold that can give rise to an event. These events must be contrasted later on with the detections performed by other telescopes in order to check which of them are really produced by the impact of meteoroids on the lunar surface. 

It must be taken into account that some regions in the images should be ignored during the impact flashes identification process in order to prevent a large number of false detections. Thus, star scintillation and also the information included frame by frame by the GPS time inserter produce sudden changes in brightness that can be interpreted by the software as impact flashes. To avoid this, a mask must be defined to indicate that regions on the image containing stars should be ignored, and just the area covered by the lunar disk should be considered. In this context, a mask is an image with the same size (width and height) as the video frames to be analyzed that specifies, by means of a colour code, which areas are going to be taken into account during the detection process and which of them will be ignored. This mask also excludes from the impact flashes detection process the area containing the information from the time inserter.

When an event \textbf{is} detected, the software creates a small AVI video file with the corresponding images and stores this file on the hard disk together with some basic information, including appearance time and the (x, y) coordinates of the centroid of the flash. Since in general these flashes are dim, the event is framed with a red square in the AVI file to allow for a better location and visualization. No brightness information is generated at this stage, as the photometric analysis is manually performed later on.

In general, under optimal atmospheric transparency, the sensitivity of our CCD video cameras makes lunar terrain features easily identifiable. These features can be used to calibrate the lunar disk, so that (x, y) coordinates in the images can be converted into selenographic coordinates (latitude and longitude on the Moon surface). MIDAS performs this calibration by means of an implemented tool that takes into account the known position of different features on the Moon's surface. In this way, we can determine the selenographic coordinates for each event recorded by the software.

On the other hand, to perform the photometric analysis of the impact flashes the brightness of each event is first obtained in device units (pixel value). The analysis is performed within a box located around the flash. The size of this box, in pixels, is specified by the user.  The same procedure is employed for reference stars, whose visual magnitude is known. Thus, by comparing the result obtained for the reference stars with that of the flash, the visual magnitude of the impact flash is inferred.
\section{Observations and results}
In order to determine which area of the lunar disk should be monitored to increase the chance of detecting impact flashes produced by Perseid meteoroids, we employed the MIDAS software. In 2012 the impact geometry of Perseid meteoroids was very unfavourable, since particles from this stream mainly hit the far side of the Moon (Figure 1a). Despite this, an observing campaign was organized during the waning crescent Moon between August 12d01h30m and August 14d05h00m UTC, with the Moon age ranging between 24.21 and 26.18 days and the disk illumination from 29 to 13 $\%$, respectively. This corresponded to a total observing time of about 8.5 hours. The Moon was covered by clouds for about 20 $\%$ of this time, and bad weather did not allow us to extend this monitoring before August 12 or after August 14. Thus, the effective observing window during this campaign corresponded to 6.9 hours. On the other hand, the peak activity of the 2012 Perseids on Earth was predicted to occur on August 12, between 12h00m and 14h30m UTC with a zenithal hourly rate (ZHR) of $\sim$ 100 meteors/hour (International Meteor Organization IMO website: http://www.imo.net). This maximum took place under daylight conditions and could not be covered by our impact flashes monitoring system.  The total area monitored by each telescope was calculated with the MIDAS software. This area was (4.1 $\pm$ 0.4)·10$^6$ km$^2$ for the 36 cm SC telescopes and (7.7 $\pm$ 0.7)·10$^6$ km$^2$ for the 28 cm SC telescope. The 40 cm newtonian telescope at La Hita Astronomical Observatory was not employed, since this system was not completely operational when the 2012 Perseid campaign was organized.
\begin{figure}
  \resizebox{\hsize}{!}{\includegraphics{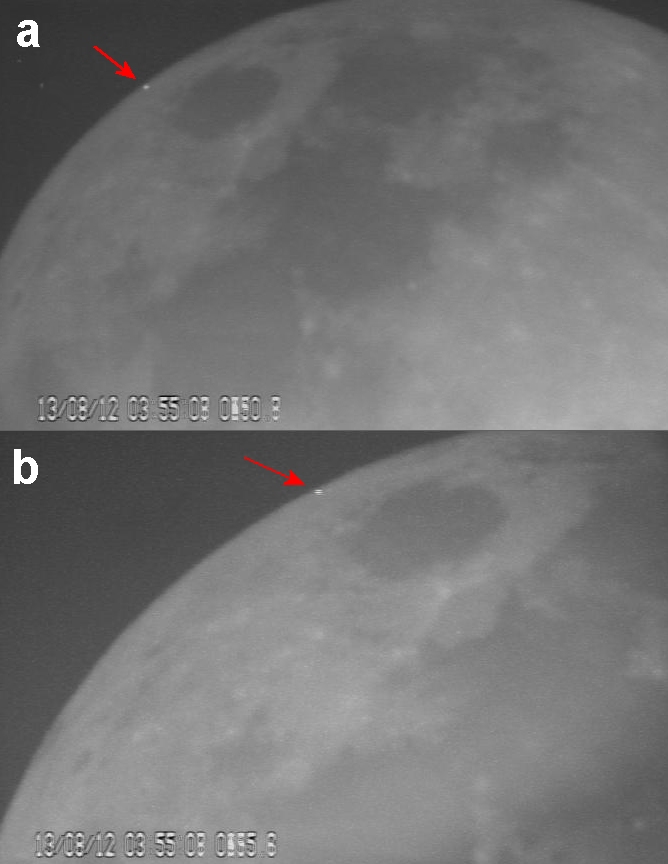}}
  \caption{Impact flash detected from Sevilla on 2012 August 13 at 3h55m08s UTC as recorded by the 0.28 cm SC telescope (a) and by one of the 0.36 cm SC telescopes (b) operating at this observatory.}
  \label{figure:2}
\end{figure}
In 2013, however, the situation was far more favourable, since the observing campaign was conducted under clear skies and the impact geometry was much more suitable (Figure 1b). This time our monitoring was developed during the waxing crescent Moon from August 11d20h02m UTC to August 13d22h20m UTC (Moon age between 4.75 and 6.88 days and disk illumination between 23 and 44 $\%$, respectively). The total (and effective) observing time was of around 6.4 hours. The area monitored on the Moon by each telescope was (5.6 $\pm$ 0.5)·10$^6$ km$^2$ (36 cm SC),  (7.4 $\pm$ 0.7)·10$^6$ km$^2$ (28 cm SC) and (5.1 $\pm$ 0.5)·10$^6$ km$^2$ (40 cm Newtonian). In this case, the peak activity of the Perseids was expected to take place on August 12, between 18h15m and 20h45m UTC with a ZHR of $\sim$ 100 meteors/hour (IMO website). 
\begin{table*}
\caption{Characteristics of the impact flashes confirmed during our 2012 and 2013 monitoring campaigns. $\tau$: flash duration; m: peak visual magnitude of the flash; $\phi$: impact angle with respect to the vertical; E$_{min}$: minimum impactor kinetic energy for flash detectability; $\nu$$^{SPO}$, $\nu$$^{PER}$: parameter $\nu$ defined by Eq. (12) for sporadics and Perseids, respectively; p$^{PER}$ probability that the impact is associated to the Perseid meteoroid stream. A luminous efficiency $\eta$=2·10$^{-3}$ has been considered}              % title of Table
\label{table:1}      % is used to refer this table in the text
\centering                                      % used for centering table
\begin{tabular}{c c c c c c c c c c}          % centered columns (5 columns)
\hline\hline                        % inserts double horizontal lines
Flash \# & Date and & Selenographic & m & $\tau$ & $\phi$ & E$_{min}$·10$^{-6}$ & $\nu$$^{SPO}$·10$^{-5}$ & $\nu$$^{PER}$·10$^{-4}$ & p$^{PER}$\\ & time (UTC) & coordinates & & (s) & (º) & (J) & & & \\   % table heading
\hline                                   % inserts single horizontal line
1 & 13/Aug/2012 & Lat: 25.2 $\pm$ 0.2 ºN & 8.2 $\pm$ 0.1 & 0.04 & 60 & 3.8 & 2.6 & 3.2 & 0.98\\
 & 03:55:08 & Lon: 83.4 $\pm$ 0.9 ºE &  &  &  &  &  &  & \\
2 & 11/Aug/2013 & Lat: 8.5 $\pm$ 0.3 ºN & 8.1 $\pm$ 0.1 & 0.04 & 73 & 3.3 & 3.1 & 3.5 & 0.96\\
 & 20:16:33 & Lon: 28.7 $\pm$ 0.3 ºW &  &  &  &  &  &  & \\ 
3 & 11/Aug/2013 & Lat: 39.4 $\pm$ 0.2 ºN & 9.1 $\pm$ 0.1 & 0.02 & 69 & 3.3 & 3.1 & 3.5 & 0.97\\
 & 20:43:19 & Lon: 49.2 $\pm$ 0.2 ºW &  &  &  &  &  &  & \\ 
4 & 11/Aug/2013 & Lat: 34.3 $\pm$ 0.3 ºN & 7.8 $\pm$ 0.1 & 0.04 & 48 & 3.3 & 3.1 & 3.5 & 0.98\\
 & 21:08:14 & Lon: 14.2 $\pm$ 0.3 ºW &  &  &  &  &  &  & \\ 
5 & 12/Aug/2013 & Lat: 17.0 $\pm$ 0.2 ºN & 6.6 $\pm$ 0.1 & 0.16 & 46 & 3.2 & 3.2 & 3.6 & 0.98\\
 & 19:49:57 & Lon: 14.6 $\pm$ 0.2 ºW &  &  &  &  &  &  & \\ 
6 & 12/Aug/2013 & Lat: 45.2 $\pm$ 0.5 ºN & 7.5 $\pm$ 0.1 & 0.08 & 25 & 3.2 & 3.2 & 3.6 & 0.99\\
 & 20:08:29 & Lon: 0.5 $\pm$ 0.5 ºW &  &  &  &  &  &  & \\ 
7 & 12/Aug/2013 & Lat: 33.4 $\pm$ 0.3 ºN & 8.8 $\pm$ 0.1 & 0.04 & 49 & 3.2 & 3.2 & 3.6 & 0.98\\
 & 20:14:55 & Lon: 28.7 $\pm$ 0.3 ºW &  &  &  &  &  &  & \\ 
8 & 13/Aug/2013 & Lat: 19.9 $\pm$ 0.2 ºN & 9.2 $\pm$ 0.1 & 0.02 & 65 & 3.2 & 3.2 & 3.6 & 0.97\\
 & 20:18:29 & Lon: 53.5 $\pm$ 0.2 ºW &  &  &  &  &  &  & \\ 
9 & 13/Aug/2013 & Lat: 50.6 $\pm$ 0.2 ºN & 9.3 $\pm$ 0.1 & 0.02 & 51 & 3.2 & 3.3 & 3.6 & 0.98\\
 & 20:39:01 & Lon: 53.6 $\pm$ 0.2 ºW &  &  &  &  &  &  & \\ 
10 & 13/Aug/2013 & Lat: 58.6 $\pm$ 0.2 ºN & 7.6 $\pm$ 0.1 & 0.04 & 39 & 3.2 & 3.3 & 3.6 & 0.98\\
 & 21:08:52 & Lon: 34.4 $\pm$ 0.2 ºW &  &  &  &  &  &  & \\ 
11 & 13/Aug/2013 & Lat: 11.3 $\pm$ 0.3 ºN & 8.5 $\pm$ 0.1 & 0.02 & 53 & 3.2 & 3.3 & 3.6 & 0.98\\
 & 21:12:47 & Lon: 32.9 $\pm$ 0.3 ºW &  &  &  &  &  &  & \\ 
12 & 13/Aug/2013 & Lat: 17.2 $\pm$ 0.2 ºN & 9.2 $\pm$ 0.1 & 0.02 & 51 & 3.2 & 3.3 & 3.6 & 0.98\\
 & 21:14:53 & Lon: 34.2 $\pm$ 0.2 ºW &  &  &  &  &  &  & \\ 
13 & 13/Aug/2013 & Lat: 63.3 $\pm$ 0.5 ºN & 7.1 $\pm$ 0.1 & 0.10 & 52 & 3.2 & 3.3 & 3.6 & 0.98\\
 & 22:03:48 & Lon: 64.6 $\pm$ 0.9 ºW &  &  &  &  &  &  & \\ 
\hline                                             %inserts single line
\end{tabular}
\end{table*}
\begin{figure}
  \resizebox{\hsize}{!}{\includegraphics{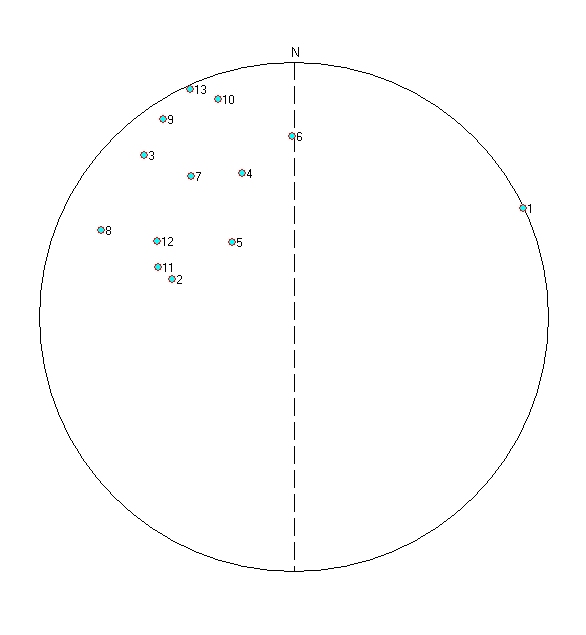}}
  \caption{Position on the lunar disk of the impact flashes identified and confirmed during the 2012 and 2013 Perseid campaigns.}
  \label{figure:3}
\end{figure}
To analyze the results obtained by our telescopes, we first discarded false detections produced by cosmic rays and electronic noise. Then, from the rest of candidates which were simultaneously detected from at least two telescopes, we found that six detections were caused by glints from artificial objects in space. Thus, for each of these six candidates, the position of the flash on the lunar disk was clearly different when the observations performed from La Hita and Sevilla were compared. Three of these six candidates lasted just one video frame. From the other three, which lasted more than one video frame, two exhibited no apparent displacement on the lunar disk. As a result of this observational effort, just one impact flash was detected and confirmed during the 2012 campaign. This event took place on 2012 August 13 at 3h55m08s UTC (Figure 2). The selenographic coordinates of the impact were 25.2 $\pm$ 0.2 ºN, 83.4 $\pm$ 0.9 ºE. During the 2013 campaign the number of confirmed impact flashes was 12. The events recorded during both campaigns are listed in Table 1 and plotted on Figure 3, and correspond to a total effective observing time of 13.3 hours.

As can be noticed, the flashes listed in Table 1 are short in duration (between 0.02 and 0.16 seconds). As explained in section 2, the visual magnitude of the impact flashes was inferred from the photometric analysis of the images. This was performed with the MIDAS software, and the calculations were double-checked with the Limovie software (Miyashita et al. 2006). A 10x10 pixels box around the flashes was employed for this photometric analysis. The resulting magnitudes are also included in Table 1. These range between 9.3 and 6.6. The longest event recorded during the Perseid campaigns considered here lasted about 0.16 seconds and took place on 2013 August 12 at 19h49m57s UTC (Figure 4). The light curve of this event, which had a peak magnitude of 6.6 $\pm$ 0.1, is shown in Figure 5. 
\begin{figure}
  \resizebox{\hsize}{!}{\includegraphics{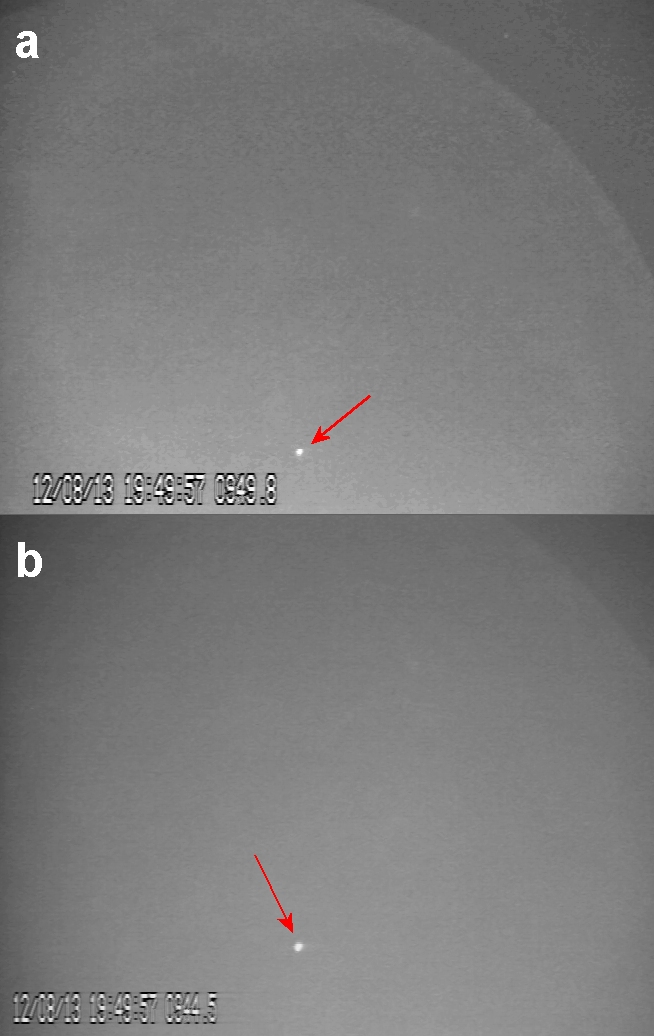}}
  \caption{Impact flash detected from Sevilla on 2013 August 12 at 19h49m57s UTC as recorded by the 0.28 cm SC telescope (a) and by one of the 0.36 cm SC telescopes (b) operating at this observatory.}
  \label{figure:4}
\end{figure}
\begin{figure}
  \resizebox{\hsize}{!}{\includegraphics{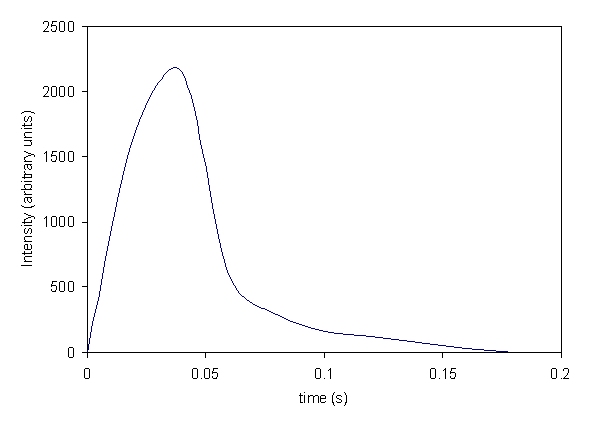}}
  \caption{Lightcurve of the impact flash recorded on 2013 August 12 at 19h49m58s UTC.}
  \label{figure:5}
\end{figure}
\section{Discussion}
\subsection{Meteoroid source}
Associating a lunar impact flash to a given meteoroid stream or to the sporadic meteoroid background is not trivial. Thus, the monitoring of these flashes provides the position vector of these collisions on the lunar surface, but not the velocity vector of the impactor. So, the orbital parameters of the meteoroid and the radiant position remain unknown. The situation is totally different when meteoroids ablate in the Earth's atmosphere: if the meteor produced by the interaction between the meteoroid and the air is recorded from at least two different locations, the velocity vector of the impactor can be calculated, but also the radiant position and the orbital elements of the particle can be easily determined (Ceplecha 1987). Once these orbital elements are known, dissimilarity criteria can be employed to test the potential association between the inferred heliocentric orbit of the meteoroid and that of a given meteoroid stream (see, e.g., Williams 2011; Madiedo et al. 2013; Madiedo 2014a, 2014b, 2015). However, despite this procedure cannot be employed for Moon impact flashes, it is still possible to establish the likely source of these flashes by correlating the results from the lunar impact monitoring with the data obtained from the monitoring of meteor activity on Earth. For this purpose we can employ the probability parameter p defined in this way:
\begin{equation}
p^{ST}=\frac{N^{ST}}{N^{ST}+N^{OTHER}},
\end{equation}
where p$^{ST}$ is the probability (ranging between 0 and 1) that the flash is associated with a given meteoroid stream, N$^{ST}$ is the number of impacts per unit time that can be produced by that stream, and N$^{OTHER}$ the number of impacts per unit time that can be produced by the rest of available sources (other streams and the sporadic background) at the same location on the lunar surface. Of course N$^{ST}$ = 0 for those flashes located out of the area on the Moon that can be impacted by meteoroids from that stream. If only \textbf{one} meteoroid stream can significantly contribute to the impact flux, N$^{OTHER}$ can be substituted by N$^{SPO}$: the number of impacts per unit time produced by the sporadic background:
\begin{equation}
p^{ST}=\frac{N^{ST}}{N^{ST}+N^{SPO}}.
\end{equation}
For simplicity we will first assume that only one meteoroid stream is active. In this case, the probability that the flash is produced by a sporadic meteoroid is
\begin{equation}
p^{SPO}=1-p^{ST}=\frac{N^{SPO}}{N^{ST}+N^{SPO}}.
\end{equation}
The flash can then be associated to the source that provides the highest value for p. 
N$^{ST}$ is related to the zenithal hourly rate (ZHR) of members of this stream impacting the Moon by means of the following equation:
\begin{equation}
N^{ST}=cos(\phi)ZHR^{ST}_{Moon},
\end{equation}
where cos($\phi$) is a geometric factor that accounts for the fact N depends on the angular distance $\phi$ from the location of the impact flash and the subradiant point (i.e., the impact angle with respect to the vertical). With respect to the sporadic background, since sporadic events come from diffuse sources these meteoroids may impact at any position on the lunar surface. For this reason hourly rates (HR) should be employed instead of zenithal hourly rates for sporadics, since the dependence of the hourly rate on the average zenithal radiant distance would be very weak (Dubietis and Artl 2010):
\begin{equation}
N^{SPO}=HR^{SPO}_{Moon}.
\end{equation}
The fact that our planet is sweeping all the dust in front of its path might have also an influence on the Moon: during the last quarter of the Moon (when the Moon is located towards the Earth's apex), the number of sporadic meteors hitting the lunar surface visible from Earth might be much lower than 15 days earlier/later when the Moon is basically trailing our planet. This effect, which would introduce an additional correction factor, has almost no influence on the results presented here and so it has not been taken into account.

On the other hand, meteoroid rates on the Moon can be related to rates on Earth by taking into account the different gravitational focusing effect between both bodies. The gravitational focusing factor $\Phi$ is given by
\begin{equation}
\Phi=1+\frac{V^{2}_{esc}}{V^{2}}.
\end{equation}
where V$_{esc}$ is the escape velocity of the central body and V the meteoroid velocity. The different gravitational focusing effect for Moon and Earth is given by the quotient $\gamma$ between the gravitational focusing factors for both bodies. For sporadic meteoroids, with an average velocity of 20 km s$^{-1}$ (Brown et al. 2002), this velocity-dependent focusing effect is higher for the Earth by a factor of 1.3 (Ortiz et al. 2006), and so $\gamma$$_{SPO}$ = 0.77. According to this we have:
\begin{equation}
HR^{SPO}_{Moon}=\gamma^{SPO}HR^{SPO}_{Earth}.
\end{equation}
\begin{equation}
ZHR^{ST}_{Moon}=\sigma\gamma^{ST}ZHR^{ST}_{Earth}.
\end{equation}
For the average hourly rate of sporadic events we have  HR$^{SPO}$$_{Earth}$=10 meteors h$^{-1}$ (Dubietis and Artl 2010). In Eq. (8) an additional factor $\sigma$ is included to take into account that the distances of Earth and Moon to the meteoric filament will in general be different, and this would give rise to a different density of stream meteoroids for both bodies. If we assume a simple situation where this filament can be approximated as a tube where the meteoroid density decreases linearly from its central axis, the following definition can be adopted for $\sigma$:
\begin{equation}
\sigma=\frac{d_{Earth}}{d_{Moon}},
\end{equation}
where d$_{Earth}$ and d$_{Moon}$ are the distance from the center of the meteoric tube to the Earth and the Moon, respectively.
On the other hand, the ZHR on Earth at solar longitude $\lambda$ (which corresponds to the time of detection of the impact flash) can be related to the peak ZHR by means of (Jenniskens 1994):
\begin{equation}
ZHR^{ST}_{Earth}=ZHR^{ST}_{Earth}(max)10^{-b|\lambda-\lambda_{max}|}.
\end{equation}
where ZHR$^{ST}$$_{Earth}$(max) is the peak ZHR on Earth (corresponding to the date given by the solar longitude $\lambda$$_{max}$). The values for the peak ZHR for different meteoroid streams and the corresponding solar longitudes for these maxima can be obtained, for instance, from (Jenniskens 2006). For streams with non-symmetrical ascending and descending activity profiles or with several maxima, Eq. (10) should be modified according to the expressions given in (Jenniskens 1994). By putting all these pieces together in Eq. (2), we can write the following expression for the probability parameter:
\begin{equation}
p^{ST}=\frac{\gamma^{ST}cos(\phi)\sigma ZHR^{ST}_{Earth}(max)10^{-b|\lambda-\lambda_{max}|}}{\gamma^{SPO}HR^{SPO}_{Earth}+\gamma^{ST}cos(\phi)\sigma ZHR^{ST}_{Earth}(max)10^{-b|\lambda-\lambda_{max}|}}
\end{equation}
However, this formula does not take into account the fundamental fact that only those meteoroids capable of producing detectable impact flashes from Earth should be included in the computations. In fact, by employing only the hourly rates measured on Earth in Eq. (11), which measures the flux of meteor brighter than mag. +6.5, it is implicitly assumed that meteoroids producing meteor events on Earth can also produce detectable impact flashes on the Moon. However, this assumption is incorrect. Thus, for a given meteoroid stream (i.e., for a given meteoroid geocentric velocity), the mass m$_o$ of meteoroids giving rise to mag. +6.5 meteors on Earth can be obtained from Eqs. (1) and (2) in Hughes (1987). For instance, this mass yields 5.0·10$^{-8}$ kg for Perseid meteoroids (V$_g$ = 59 km s$^{-1}$), 2.4·10$^{-8}$ kg for Leonids (V$_g$ = 70 km s$^{-1}$) and 5.0·10$^{-6}$ kg for sporadic meteoroids with an average velocity of 20 km s$^{-1}$ (Brown et al. 2002). However, the masses corresponding to impact flashes recorded on the Moon are several orders of magnitudes larger than m$_o$ (see, e.g., Ortiz et al. 2006, Yanagisawa et al. 2006, Swift et al. 2011). So, the minimum kinetic energy E$_{min}$ to produce a detectable impact flash on the Moon is much higher than the kinetic energy of a portion of the meteoroids included in the computation of hourly rates on Earth. This means that the velocity and mass distribution of meteoroids must be somehow included in Eq. (11) to take into consideration for the computation of the probability parameter p only those meteoroids with a kinetic energy above the threshold kinetic energy given by E$_{min}$ (the method for the computation of E$_{min}$ will be explained below). According to Eq. (2) in Bellot Rubio et al. (2000), this can be accomplished by including in Eqs. (2) and (11) the following factor:
\begin{equation}
\nu=\left(\frac{m_{o}V^{2}}{2}\right)^{s-1}E^{1-s}_{min},
\end{equation}
where V is the impact velocity, m$_o$ is the mass of a shower meteoroid producing on Earth a meteor of magnitude +6.5 and s is the mass index, which is related to the population index r (the ratio of the number of meteors with magnitude m+1 or less to the number of meteors with magnitude m or less) by means of the relationship:
\begin{equation}
s=1+2.5log(r).
\end{equation}
According to the definition of $\nu$, this parameter is different for each meteoroid stream (and for sporadic meteoroids, of course). By taking this into account Eq. (11) should be modified as follows:
\begin{equation}
p^{ST}=\frac{\nu^{ST}\gamma^{ST}cos(\phi)\sigma ZHR^{ST}_{Earth}(max)10^{-b|\lambda-\lambda_{max}|}}{\splitfrac{\nu^{SPO}\gamma^{SPO}HR^{SPO}_{Earth}}{+\nu^{ST}\gamma^{ST}cos(\phi)\sigma ZHR^{ST}_{Earth}(max)10^{-b|\lambda-\lambda_{max}|}}}
\end{equation}
If by the time of detection of the impact flash n additional meteoroid streams with significant contribution to the impact rate (and with compatible impact geometry) must be considered, the denominator in Eq. (14) must be modified in the following way:
\begin{equation}
p^{ST}=\frac{\nu^{ST}\gamma^{ST}cos(\phi)\sigma ZHR^{ST}_{Earth}(max)10^{-b|\lambda-\lambda_{max}|}}{\splitfrac{\nu^{SPO}\gamma^{SPO}HR^{SPO}_{Earth}}{+\nu^{ST}\gamma^{ST}cos(\phi)\sigma ZHR^{ST}_{Earth}(max)10^{-b|\lambda-\lambda_{max}|}+\kappa}}
\end{equation}
where 
\begin{equation}
\kappa=\sum_{i=1}^{n}\nu^{ST}_i\gamma^{ST}_i cos(\phi_i)\sigma ZHR^{ST}_{i,Earth}(max)10^{-b_i|\lambda_i-\lambda_{i,max}|}
\end{equation}
accounts for these n additional streams.
The minimum kinetic energy E$_{min}$ defined above corresponds to the minimum radiated energy E$_{r\_min}$ on the Moon detectable from observations on Earth, which in turn is related to the maximum visual magnitude for detectable impacts (m$_{max}$). And these values depend, among other factors, on the experimental setup employed. The kinetic energy of the impactor and the radiated energy are linked by the luminous efficiency:
\begin{equation}
E_{r\_min}=\eta E_{min}.
\end{equation}
With our experimental setup, the maximum visual magnitude for detectable impact is m$_{max}$ $\sim$ 10. The radiated energy can be obtained by integrating the radiated power P defined by this equation:
\begin{equation}
P=1.36949\times 10^{-16}10^{(-m+21.1)/2.5}f \pi \Delta \lambda R^2.
\end{equation}
where P is given in Joules, m is the magnitude of the flash, 1.36949·10$^{-16}$ is the flux density in W m$^{-2}$ $\mu$m$^{-1}$ for a magnitude 21.1 source according to the values given in Bessel (1979), $\Delta$$\lambda$ is the width of the filter passband (about 6000 \AA\ for our devices) and R is the Earth-Moon distance at the instant of the meteoroid impact. The factor f is related to the degree of anisotropy of light emission. Thus, for those impacts where light is isotropically emitted from the surface of the Moon f = 2, while f = 4 if light is emitted from a very high altitude above the lunar surface. As expected, according to Eq. (18) the minimum radiated power for flash detectability (and hence the minimum radiated energy and the minimum meteoroid kinetic energy), which is obtained by using m = m$_{max}$, is higher the larger is the distance between Earth and Moon. So, the detectability limit is time-dependent.

In our analysis we have considered that the events recorded during our 2012 and 2013 Perseid campaigns could be produced either by Perseid meteoroids or by particles from the sporadic background. The corresponding values of the probability parameter for Perseids (p$^{PER}$) obtained from equation (14) for these impact flashes are listed in Table 1. For this calculation the angle $\phi$ was determined by the MIDAS software and the value of b (0.20) was taken from (Jenniskens 1994). For simplicity we have considered $\sigma$ $\sim$ 1. For the average hourly rate of sporadic events we have taken  HR$^{SPO}$$_{Earth}$=10 meteors h$^{-1}$ (Dubietis and Artl 2010). To calculate the values of $\nu$$^{PER}$ and $\nu$$^{SPO}$ from Eq. (12) we have employed the typical value r = 2.0 for the Perseids (Brown and Rendtel 1996), while for sporadic meteors the average value r = 3.0 has been adopted (see e.g. Dubietis and Artl 2010; Rendtel 2006). An average value of 17 km s$^{-1}$ has been assumed for the impact velocity on the Moon of sporadic events (Ortiz et al. 1999). However, for Perseid meteoroids the impact velocity depends on the impact geometry and must be determined follows. First, from the known position of the radiant and the modulus of the geocentric velocity (V$_g$ = 59 km s$^{-1}$), we calculated the geocentric velocity vector $\Vec{V}$$_g$. Next, the heliocentric velocity of Earth $\Vec{V}$$_E$ was obtained by means of the JPL Horizons online ephemeris generator (http://ssd.jpl.nasa.gov/?horizons). From both vectors we calculated the heliocentric velocity vector $\Vec{V}$$_h$ of the meteoroid:
\begin{equation}
\Vec{V}_h=\Vec{V}_E+\Vec{V}_g.
\end{equation}
On the other hand, the heliocentric velocity of the Moon $\Vec{V}$$_M$ was also obtained from the JPL Horizons online ephemeris generator. The selenocentric velocity vector $\Vec{V}$$_S$ of the meteoroid is given by:
\begin{equation}
\Vec{V}_S=\Vec{V}_h-\Vec{V}_M.
\end{equation}
and from the modulus of $\Vec{V}$$_S$ we inferred the modulus of the impact velocity V by adding the escape velocity of the Moon V$_{EM}$ (2.4 km s$^{-1}$):
\begin{equation}
V^2=V_S^2+V_{EM}^2.
\end{equation}
Nevertheless, in the cases considered here we have found that the value of V obtained for each meteoroid by following this procedure differs from V$_g$ by at most 0.1 km s$^{-1}$, which is below the accuracy of the value taken for V$_g$. Thus, we have considered V = 59 km s$^{-1}$. The minimum kinetic energy E$_{min}$ was obtained from Eqs. (17) and (18) by considering f = 2 and $\eta$ = 2·10$^{-3}$ (Ortiz et al. 2006). This parameter, together with the values inferred for $\nu$$^{PER}$, $\nu$$^{SPO}$ and p$^{PER}$, is shown in Table 1. The value of the probability parameter p$^{PER}$ is well above 0.5 for all these events, which implies that these flashes can be considered as produced by Perseid meteoroids impacting the Moon. This probability, as previously mentioned, is based on the assumption that $\eta$ = 2·10$^{-3}$. So, we have checked the stability of this result by using different values for the luminous efficiency. Smaller values of $\eta$ (up to an order of magnitude below the assumed value) gave rise to slightly higher probabilities. But the situation was the opposite for larger values of the luminous efficiency, which gave rise to lower p$^{PER}$ values. However, the probability parameter remains above 0.8 for all these flashes even when $\eta$ is two order of magnitudes larger than the assumed value, which confirms that these events can be assigned on solid grounds to Perseid meteoroid impacts.
\subsection{Luminous efficiency and impactor mass}
\begin{figure}
  \resizebox{\hsize}{!}{\includegraphics{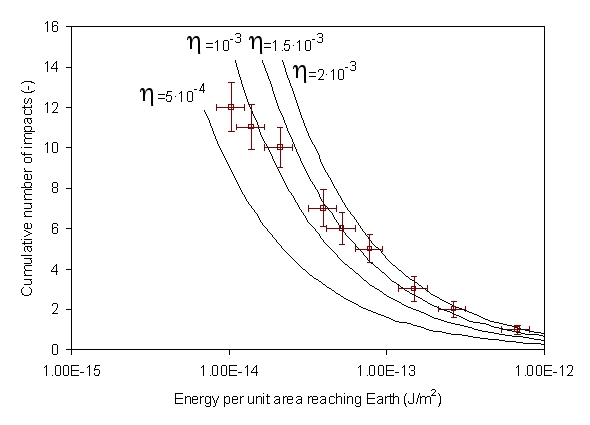}}
  \caption{Comparison between the observed (open squares) and expected (solid lines) cumulative number of impacts within the field of view of the camera for different values of the luminous efficiency.}
  \label{figure:6}
\end{figure}
Once we established that the flashes in Table 1 could be produced by the collision of Perseid meteoroids, we followed the method employed by Bellot Rubio et al. (2000a,b) to obtain the luminous efficiency for lunar impact flashes produced by meteoroids belonging to the Perseid stream in 2013. The result from the 2012 survey was not included, since just one impact flash was recorded and to employ this approach several events are necessary. According to this method, the number N of expected impact flashes above an energy E$_d$ is given by:
\begin{equation}
N(E_d)=\int_{t_o}^{t_o+\Delta t} F(m_o,t)\Delta t\left(\frac{2f\pi R^2}{\eta m_o V^2}E_d\right)^{1-s}\,A\mathrm{d}t,
\end{equation}
where t$_{o}$ and t$_{o}$ +$\Delta$t define the observing time interval, E$_d$ is the time-integrated optical energy flux of the flash observed from Earth, F(m$_{o}$,t) is the flux or meteoroids on the Moon with mass higher than m$_o$ at time t, and A is the projected area of the observed lunar surface (i.e., visible from Earth) perpendicular to the Perseid stream. In our case A = 1.2·10$^{6}$ km$^{2}$ and $\Delta$t = 6.4 h. We have assumed f = 2. As mentioned in previous sections, for Perseid meteoroids we have V = 59 km s$^{-1}$, m$_{o}$ = 5.0·10$^{-8}$ kg, and r = 2.0. This value of the population index implies that the mass index s given by Eq. (13) yields 1.7. According to the focusing factor defined by Eq. (6), the flux of meteoroids on the Earth is higher than the flux on the moon by a factor of 1.04. The dependence of this flux with time can be obtained by expressing the flux of meteoroids as a function of the solar longitude $\lambda$ (Jenniskens 1994):
\begin{equation}
F(m_o,\lambda)=F(m_o,\lambda_{max})10^{-b|\lambda-\lambda_{max}|},
\end{equation}
where F(m$_{o}$,$\lambda$$_{max}$) is the peak flux corresponding to the date given by the solar longitude $\lambda$$_{max}$. According to the observations performed by our meteor observing stations (Madiedo \& Trigo-Rodriguez 2008, Madiedo 2014), we have found that on Earth the peak flux for Perseids is 2.0·10$^{-2}$ meteors km$^{-2}$ h$^{-1}$. The value of b for this meteoroid stream (0.20) was taken from (Jenniskens 1994). The predicted value of the cumulative number of impacts calculated from Eq. (22) and the experimental value obtained for this magnitude are plotted in Figure 6. The prediction has been performed for different values of the luminous efficiency. This plot shows that for the largest impactors (those with values of E$_d$ between 10$^{-13}$ and 10$^{-12}$ J m$^{-2}$) the best fit is obtained when $\eta$ = 2·10$^{-3}$. This coincides with the value of $\eta$ obtained from the analysis of Leonid impact flashes by Bellot Rubio et al. (2000a,b) and Ortiz et al. (2002). However, for values of E$_d$ between 5·10$^{-14}$ and 10$^{-13}$ J m$^{-2}$ a luminous efficiency of about 1.5·10$^{-3}$, which coincides with the value obtained by Swift et al. (2011), provides a better match. For values of E$_{d}$ below 5·10$^{-14}$ J m$^{-2}$ the experimental data start to deviate significantly from the prediction obtained with these efficiencies. This might be due to the fact that some faint events may have been missed, and so the lower mass end of the cumulative number of impacts should be taken with care. Thus, by ignoring this mass end, we have obtained that the best fit between the predicted and experimental values is obtained for $\eta$ = 1.8·10$^{-3}$. In any event, our calculation confirms that the assumption $\eta$ = 2·10$^{-3}$ performed in the previous section to calculate the probability parameter value was realistic.
Once the luminous efficiency was calculated, we obtained the mass of the meteoroids that produced the flashes listed in Table 1. This mass was inferred from the kinetic energy E of these projectiles calculated from the radiated power given by Eq. (18). The integration of this power provided the radiated energy Er recorded by the telescopes, with E$_{r}$ = $\eta$E. Then, for the meteoroid mass M we have:
\begin{equation}
M=2EV^{-2},
\end{equation}
As Table 2 shows, these masses range from about 1.9 to 190 g when the luminous efficiency is set to $\eta$ = 1.8·10$^{-3}$. The corresponding meteoroid size varies between around 1.4 and 6.3 cm in diameter, by considering an average bulk density for Perseid meteoroids of 1.2 g cm$^{-3}$ (Babadzhanov and Kokhirova 2009).
\begin{table*}
\caption{Impactor mass M and diameter D$_p$, and crater size D, derived for the Perseid impact flashes analyzed in the text. A value of $\eta$ = 1.8·10$^{-3}$ has been considered.}              % title of Table
\label{table:2}      % is used to refer this table in the text
\centering                                      % used for centering table
\begin{tabular}{c c c c c}          % centered columns (5 columns)
\hline\hline                        % inserts double horizontal lines
Flash \# & Date and & M & D$_p$ & D\\ & time (UTC) & (g) & (cm) & (m) \\   % table heading
\hline                                   % inserts single horizontal line
1 & 13/Aug/2012 & 12.7 $\pm$ 1.1 & 2.4 $\pm$ 0.1 & 1.99 $\pm$ 0.05 \\
 & 03:55:08 &  &  &  \\
2 & 11/Aug/2013 & 12.1 $\pm$ 1.1 & 2.7 $\pm$ 0.1 & 1.80 $\pm$ 0.04 \\
 & 20:16:33 &  &  &  \\ 
3 & 11/Aug/2013 & 2.4 $\pm$ 0.2 & 1.4 $\pm$ 0.1 & 1.14 $\pm$ 0.03\\
 & 20:43:19 &  &  &  \\ 
4 & 11/Aug/2013 & 16.0 $\pm$ 1.4 & 2.7 $\pm$ 0.1 & 2.44 $\pm$ 0.06\\
 & 21:08:14 &  &  &  \\ 
5 & 12/Aug/2013 & 190 $\pm$ 17 & 6.3 $\pm$ 0.2 & 5.09 $\pm$ 0.13\\
 & 19:49:57 &  &  &  \\ 
6 & 12/Aug/2013 & 44 $\pm$ 4 & 4.0 $\pm$ 0.1 & 3.76 $\pm$ 0.08\\
 & 20:08:29 &  &  &  \\ 
7 & 12/Aug/2013 & 6.2 $\pm$ 0.5 & 2.3 $\pm$ 0.1 & 2.07 $\pm$ 0.05\\
 & 20:14:55 &  &  &  \\ 
8 & 13/Aug/2013 & 2.1 $\pm$ 0.2 & 1.4 $\pm$ 0.1 & 1.17 $\pm$ 0.04\\
 & 20:18:29 &  &  & \\ 
9 & 13/Aug/2013 & 1.9 $\pm$ 0.2 & 1.4 $\pm$ 0.1 & 1.31 $\pm$ 0.05\\
 & 20:39:01 &  &  &  \\ 
10 & 13/Aug/2013 & 18.5 $\pm$ 1.6 & 2.8 $\pm$ 0.1 & 2.64 $\pm$ 0.06\\
 & 21:08:52 &  &  &  \\ 
11 & 13/Aug/2013 & 4.0 $\pm$ 0.3 & 1.5 $\pm$ 0.1 & 1.43 $\pm$ 0.03\\
 & 21:12:47 &  &  & \\ 
12 & 13/Aug/2013 & 2.1 $\pm$ 0.2 & 1.4 $\pm$ 0.1 & 1.34 $\pm$ 0.04\\
 & 21:14:53 &  &  & \\ 
13 & 13/Aug/2013 & 73 $\pm$ 6 & 4.5 $\pm$ 0.2 & 3.74 $\pm$ 0.09\\
 & 22:03:48 &  &  & \\ 
\hline                                             %inserts single line
\end{tabular}
\end{table*}
\subsection{Crater size}
The size of the crater excavated by a meteoroid impacting on the lunar surface can be calculated from the following crater-scaling equation for the Moon given by Gault, which is valid for craters with a diameter of up to about 100 meters in loose soil or regolith (Gault 1974, Melosh 1989):
\begin{equation}
D=0.25\rho^{1/6}_{p}\rho^{-0.5}_{t}E^{0.29}(sin(\theta))^{1/3}.
\end{equation}
In this relationship, where magnitudes must be entered in mks units, D is the crater diameter, E is the kinetic energy of the impactor, $\rho$$_p$ and $\rho$$_t$ are the impactor and target bulk densities, respectively, and $\theta$ = 90º-$\phi$ is the impact angle with respect to the horizontal. For the target bulk density we have $\rho$$_t$=1.6 g cm$^{-3}$. For the bulk density of Perseid meteoroids we have assumed $\rho$$_p$=1.2 g cm$^{-3}$ (Babadzhanov and Kokhirova 2009). For each case, the impact angle with respect to the local horizontal was determined by the MIDAS software from the impact geometry. The resulting crater sizes are shown in Table 2. As can be seen, the crater diameter ranges from around 1.14 to 5.09 m. Of course, these craters are too small to be directly observed from Earth, but they could be observed by means of probes orbiting the Moon, such as Lunar Reconnaissance Orbiter (LRO).
\subsection{Impact duration}
\begin{figure}
  \resizebox{\hsize}{!}{\includegraphics{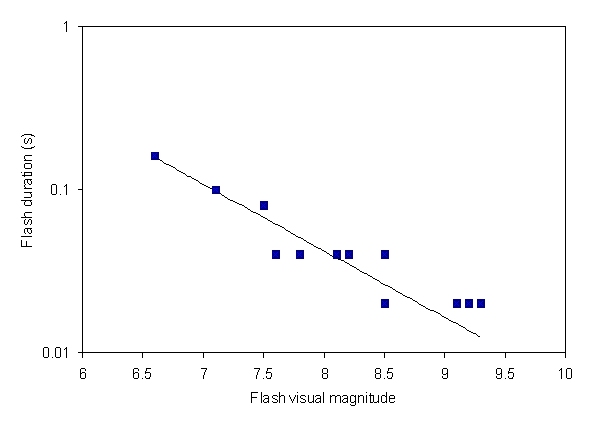}}
  \caption{Experimental (squares) and fitted (solid line) values of impact flash duration as a function of the visual magnitude of the event.}
  \label{figure:7}
\end{figure}
We have investigated the relationship between the duration of these impact flashes and their visual magnitude. With our CCD video cameras this duration can be established with an accuracy of 0.02 seconds, which corresponds to the duration of a single frame when we deinterlace the interlaced AVI video files obtained by these devices at 25 fps. As can be seen in Figure 7 there is a correlation between both quantities, since the duration ($\tau$) exhibits an exponential dependence with magnitude (m). Thus, the following relationship was found by fitting the values listed in Table 1:
\begin{equation}
\tau=(77.6 \pm 34.4)exp[(-0.94 \pm 0.06)m].
\end{equation}
This exponential dependence was also found by Bouley et al. (2012) for impact flashes associated with different sources.
\section{Conclusions}
In this paper we have presented the analysis of the lunar impact flashes recorded during the 2012 and 2013 Perseid monitoring campaigns developed in the framework of our MIDAS project. In total, 13 impact flashes were recorded and confirmed. Just one of these was identified during the 2012 monitoring campaign. The 2013 campaign, which was developed under much more favourable weather and geometric conditions, provided 12 confirmed events. The brightness of these impact flashes ranged from mag. 9.3 to 6.6. Besides, their duration exhibits an exponential dependence with the visual magnitude of these events.

On the other hand, we have developed a criterion to quantify the probability of association of impact flash with a given meteoroid stream. The relationships derived in this paper show that the events discussed here were most likely produced by Perseid meteoroids. Thus, this probability is always above 96$\%$. The value inferred for the luminous efficiency of these Perseid impact flashes is 1.8·10$^{-3}$. With this value we have obtained that the impactor mass ranged between 1.9 and 190 g. The craters produced on the lunar surface by these impacts have sizes ranging from about 1.14 to 5.09 m in diameter. Despite these fresh craters cannot be observed with instruments on Earth, they could be observed by means of probes orbiting the Moon, such as LRO.

\bibliographystyle{aa}

\end{document}